\documentclass[runningheads]{llncs}
\usepackage{tabularx}
\usepackage{diagbox}
\usepackage{multirow}
\usepackage{graphicx}
\usepackage{amsmath}
\usepackage{array}

\begin{document}
\title{The role of MRI physics in brain segmentation CNNs: achieving acquisition invariance and instructive uncertainties}
\author{
Pedro Borges\inst{1,2} \and
Richard Shaw\inst{1,2} \and
Thomas Varsavsky\inst{1,2} \and
Kerstin Klaser\inst{2}
David Thomas\inst{3} \and
Ivana Drobnjak\inst{1} \and
Sebastien Ourselin\inst{2} \and
M Jorge Cardoso\inst{2}}

\authorrunning{P. Borges et al.}
%
\institute{$^1$Department of Medical Physics and Biomedical Engineering, UCL, UK \\ $^2$School of Biomedical Engineering and Imaging Sciences, KCL, UK \\ $^3$Dementia Research Centre, UCL, UK \\
}
%
\maketitle              
\begin{abstract}
Being able to adequately process and combine data arising from different sites is crucial in neuroimaging, but is difficult, owing to site, sequence and acquisition-parameter dependent biases. It is important therefore to design algorithms that are not only robust to images of differing contrasts, but also be able to generalise well to unseen ones, with a quantifiable measure of uncertainty. In this paper we demonstrate the efficacy of a physics-informed, uncertainty-aware, segmentation network that employs augmentation-time MR simulations and homogeneous batch feature stratification to achieve acquisition invariance. We show that the proposed approach also accurately extrapolates to out-of-distribution sequence samples, providing well calibrated volumetric bounds on these. We demonstrate a significant improvement in terms of coefficients of variation, backed by uncertainty based volumetric validation.

\end{abstract}
\section{Introduction}
Magnetic Resonance Imaging (MRI) is one of the most widespread neuroimaging techniques owing to its excellent soft tissue contrast, boasting great versatility in highlighting different regions and pathologies by means of sequence selection. As a consequence, a significant body of work has emerged developing accurate processing algorithms for MR images that may arise from different sites and acquisition sequence parameters. 
There are those works that focus on achieving algorithms that can generalise well to all contrasts. Traditional and largely widespread techniques include probabilistic generative models~\cite{AshburnerFriston2005} and multi-atlas fusion methods~\cite{Sabuncu2010}. However, the former has strong assumptions on label intensity distributions, and the latter is predicated on lengthy processing times due to its dependence on image registration. Recent works using convolutional neural networks (CNNs), such as Billot \textit{et al.}~\cite{Billot2020a}, tackle contrast agnosticism by employing a Bayesian generative segmentation model that synthesises images containing multiple different contrasts. Jog \textit{et al.}~\cite{Jog2019} devise an approach by which networks can be made to generalise to unseen contrasts by predicting pulse sequence parameters from such images, and simulating images of that contrast by using labelled multiparametric map datasets. Pham \textit{et al.}~\cite{10.1007/978-3-030-59520-3_1} employ an iterative approach involving a dual segmentation-synthesis model, whereby images of unseen contrasts are segmented, used to train a synthesis network that in turn is used to generate new images of the unseen contrast from the labels in the original training set. It is important to note that, while these methods are able to segment data from unseen sites with some degree of accuracy, they do not model the interaction between acquisition parameters and the underlying anatomy explicitly - they segment what they see and not the true anatomy.

This leads to those methods that seek to harmonise measurements across sites by directly accounting for such covariates as scanner and site bias, and sequence contrast variabilities, e.g. ComBat~\cite{Johnson2007} is a Bayesian framework designed to account for experimental variabilities that has been applied to cortical thickness harmonisation~\cite{Fortin2018}. These classes of techniques, however, operate directly on extracted volumetric measurements and not on the images. Harmonisation has also been tackled with CycleGANs~\cite{Zhu2017}~\cite{Zhao2019} and domain adaptation approaches~\cite{Dinsdale2020}.

Recent work~\cite{Borges2020} proposed a means to directly introduce the physics of the MR acquisition process directly into deep learning networks in combination with pre-generated synthetic MR images based on multi-parametric MR maps (MPMs). This work achieves some agnosticism to the underlying physics by demonstrating that generated segmentations are more consistent volumetrically. This method, however, does not enforce volumetric consistency across contrasts, and has not been show to extrapolate to out of distribution sequence parameters. 

Changes in MRI acquisition parameters alter the tissue contrast, thus impacting the algorithmic ability to accurately segment images; this can be modeled via uncertainty estimation. Here, we propose to model both epistemic (ability of the model to know) and aleatoric (unknowns of the data) uncertainties. Building on existing work~\cite{Borges2020}, we also introduce a new training approach and consistency loss across realisations of MRI contrasts, allowing the model to appropriately disentangle the anatomical phenotype and the MRI physics, and extrapolate to unseen contrasts without sacrificing segmentation quality.

\section{Methods}
Borges \textit{et al.}~\cite{Borges2020} proposed that a network could be made resilient to changes in the physics parameters, and therefore be able to appropriately segment data produced by different sequences. This was achieved by generating simulated data, and passing this imaging data and associated MRI parameters to a CNN. In order to train against a "Physics Gold Standard" (PGS), i.e. a true model of the anatomy that is not influenced by the choice of acquisition parameters, the authors used a Gaussian Mixture Model of literature sourced tissue parameters for grey matter (GM), white matter (WM), and cerebrospinal fluid (CSF) on their quantitative MPMs. We build on this work and improve the algorithmic robustness, ameliorating image quality, segmentation volume consistency, and validating within and out of distribution samples paired with uncertainty derived errors.

\begin{figure}[t!]
\centering
\includegraphics[width=1.\textwidth]{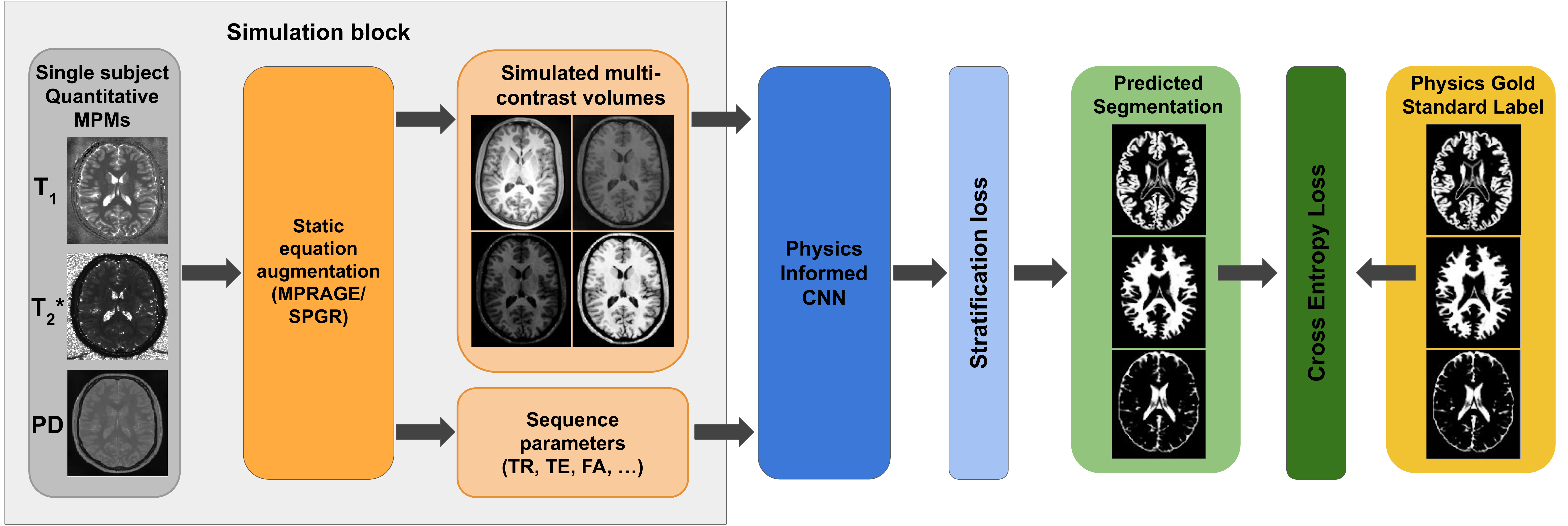} 
\caption{The training pipeline with proposed new additions of single subject batch stratification and accompanying $L_2$ feature maps loss, and training time image simulation.}
\label{fig:figPipeline}
\end{figure}

\subsection{Network architecture}
In Borges \textit{et al.}~\cite{Borges2020}, the injection of the physics parameters into the network is done via the inclusion of two fully connected layers whose output is tiled and concatenated to the ante-penultimate convolutional layer output. We adopt a similar strategy, but take the added step of also tiling this output to an earlier region of the network, immediately preceding the first down-sampling layer. We argue that knowledge of the physics is potentially valuable information in the encoding portion of the network, and that this allows it to better disentangle the physics parameters and the subject's phenotype.

 We moved to adopt the nn-UNet architecture~\cite{Isensee2018}. All networks were trained with batch size 4, on 3D patches of size $128^3$ sampled from the simulated volumes. Networks were trained with a learning rate of $10^{-4}$ until convergence, where convergence is defined as 7 epochs elapsing without an improvement in validation metrics, Dice score combined with coefficient of variation (CoV). We made use of two main frameworks for this work, TorchIO ~\cite{Perez-Garcia2020}, and MONAI ~\cite{MONAI}.

As the proposed method requires multi-parametric data to train the model, a more scarce resource in large numbers, a dataset comprised of 18 subjects were used for training, four for validation, and five for inference/ testing.

\subsection{Stratification and batch homogeneity}
We seek to further enforce volumetric consistencies vis-à-vis same-subject realisations generated using different sequence parameters. We therefore propose a batch stratification approach where each batch contains multiple realisations of images from a single subject. This allows for the addition of a stratification loss over the batch features of the penultimate layer of our network, which acts in addition to the standard cross-entropy segmentation loss. 
As the segmentation ground truths remain consistent across same subject simulations (because the underlying anatomy is unchanging), if a given batch contains multiple simulations from a single subject (and same patch location for patch-based training networks), then the features maps at the end of the network should also be consistent across simulations. This is enforced by introducing an $L_2$ loss over all the final feature maps for each batch. The inclusion of the physics parameters should make this tenable, as it allows for the network to learn to disentangle the anatomical phenotype and the MRI-physics related appearance. 

\subsection{Casting simulation as an augmentation layer}
We adopt the same static equation multi-parametric map based simulation approach as Jog \textit{et al.}~\cite{JOG}, focusing on MPRAGE and SPGR sequences. The SPGR equation describing the signal $b_S$ per voxel, $x$ is: 

\begin{equation} \label{eq:eqSPGR}
b_S(x) = G_SPD(x)sin{\theta}\frac{1-e^{-\frac{TR}{T_1(x)}}}{1-\cos{\theta}e^{-\frac{TR}{T_1(x)}}}e^{-\frac{TE}{T_2^{*}(x)}},
\end{equation}
where $G_S$ is the scanner gain, $TR$ the repetition time, $T_1$ the longitudinal relaxation time, $TE$ the echo time, and $T_2^{*}$ the transverse relaxation time.

Similarly, for MPRAGE:

\begin{equation} \label{eq:eqMPRAGE}
b_M(x) = G_SPD(x)\Bigg(1-\frac{2e^{\frac{-TI}{T_1(x)}}}{1+e^{\frac{-(TI+TD+\tau)}{T_1(x)}}}\Bigg),
\end{equation}
\newline
where $TD$ the delay time, and $\tau$ the slice imaging time.

Unlike in ~\cite{Borges2020}, where simulated volumes are all generated prior to network training, we implemented the static equation simulation layer as an augmentation layer. Such a layer takes as input a 4D multi-parametric map, a protocol type, and a range of relevant parameters to randomly sample from, producing N (batch size) simulated volumes. This layer-based batch approach is compatible with our posed stratification model, as all generated volumes per batch belong to the same subject, permitting the utilisation of the within-batch feature consistency loss. The full training pipeline is depicted in Fig.~\ref{fig:figPipeline}.

\subsection{Uncertainty modelling}
We opt to incorporate uncertainty modelling in our framework to obtain volumetric bounds on our segmentations. We model the aleatoric uncertainty via explicit loss attenuation~\cite{Kendall}. We modify our network architecture to include an additional convolutional block that branches off the final upsampling layer. This branch models the aleatoric uncertainty, $\sigma_{Het}^W(x)$. This modifies the cross-entropy loss function accordingly:
\begin{align} \label{eq:eqHeteroLoss}
\hat{x}_{i,t} = f_i^W + \epsilon_t, \qquad \epsilon_t \sim \mathcal{N}(0, (\sigma_i^W)^2) \\
\mathcal{L}=\sum_{i}log\frac{1}{T}\sum_{t}c_{i}(-\hat{x}_{i,t,c} + log\sum_{c'}e^{\hat{x}_{i,t,c'}}) 
\end{align}
Where $\hat{x}_{i,t}$ are the task logits ($f_i^W$) summed with a noise sample of standard deviation equal to the predicted $\sigma_{i}^W$ per voxel; $T$ denotes the number of stochastic passes per input, and $\sigma^W$ is defined for every voxel, per class, $c$. This allows for the easy extraction of volumetric bounds by repeatedly sampling from additive logit noise distributions to produce new segmentations.

The epistemic uncertainty is modelled using test-time Monte Carlo sampling via dropout. Dropout is commonly used as a regularisation technique~\cite{Srivastava2014}, but also allows for the approximate Bayesian posterior sampling of segmentations by maintaining the random neuron switching at test-time~\cite{Gal2015}. We set a dropout level of 0.5 in all layers except for the input layer, where it is set to 0.05.

\section{Experiments}

\subsection{Data}
We make use of a 27 subject multi-parametric early onset Alzheimer dataset, the same as in~\cite{Borges2020}, for the purpose of simulating images which are used for training, validating, and testing of our models, all of which are registered to MNI space rigidly. The images contain maps of the longitudinal and effective transverse magnetisation relaxation, $R_1$ and $R_2^{*}$, proton density, $PD$, and magnetisation transfer, MT. The details concerning quantitative map creation can be found in~\cite{helms2009increased}. The static equation models we employ feature $T_1$ (inverse of $R_1$), $T_2^{*}$ (inverse of $R_2^{*}$), and PD.

\subsection{Simulation sequence details}
To allow for direct comparability, we limited the ranges of the relevant parameters for simulated images at training time to those stipulated in the original work, i.e. inversion time (TI) = [600-1200] ms for MPRAGE, repetition time (TR) = [15-100] ms, echo time (TE) = [4-10] ms, and flip angle (FA) = [15-75] degrees for SPGR. For each subject, a single "Physics Gold Standard" (PGS) segmentation was used across the associated synthesized images, generated using the same process and literature values as in the original work~\cite{Borges2020}.

\section{Annealing study: Robustness and quality analysis}
To ascertain the contributions of the two main additions to the underlying method, we carry out an annealing study, whereby we analyse the incremental performance increases in terms of volume consistency and Dice score, with the addition of each change. We begin with a complete physics-agnostic baseline, i.e. a standard 3D nn-UNet trained with pre-generated data (Baseline), followed by the original physics method (Phys-Base), followed by Phys-Base with the addition of batch stratification (Phys-Strat), followed lastly by Phys-Strat with the addition of the simulation augmentation scheme (Phys-Strat-Aug).

\newcolumntype{d}{X}
\newcolumntype{s}{>{\hsize=.5\hsize}X}
\newcommand{\heading}[1]{\multicolumn{1}{c}{#1}}
\newcolumntype{P}[1]{>{\centering\arraybackslash}p{#1}}
\newcolumntype{Y}{>{\centering\arraybackslash}X}
\newcolumntype{v}{>{\hsize=.5\hsize}Y}

\begin{table}[t!]
\caption{Mean dice scores for Baseline, Phys-Base, Phys-Strat, and Phys-Strat-Aug on segmentation task, across inference subjects. All dice scores are estimated against a Physics Gold Standard. Standard deviations quoted in brackets.  Bold values represent statistically best performances.}\label{tab:tabDice}
\centering
\begin{tabularx}{\linewidth}{|Y|v|v|v|v|v|v|v|v|}
\hline
\multirow{3}{*}{Experiments} & \multicolumn{8}{c|}{Sequence Dice Scores} \\ \cline{2-9}
& \multicolumn{4}{c|}{MPRAGE} & \multicolumn{4}{c|}{SPGR} \\ \cline{2-9}
& \multicolumn{2}{c|}{GM} & \multicolumn{2}{c|}{WM} & \multicolumn{2}{c|}{GM} & \multicolumn{2}{c|}{WM} \\ \cline{2-9}
& IoD & OoD & IoD & OoD & IoD & OoD & IoD & OoD \\
\hline

\multirow{2}{*}{\parbox{1.5cm}{Baseline}} &
0.966 & 0.956 & 0.953 & 0.934 &
0.878 & 0.872 & 0.893 & 0.873 \\
& (0.005) & (0.006) & (0.002) & (0.002) &
(0.021) & (0.008) & (0.023) & (0.011) \\

\hline

\multirow{2}{*}{\parbox{1.5cm}{Phys-Base}} &
\textbf{0.971} & 0.964 & \textbf{0.964} & \textbf{0.959} &
0.911 & 0.872 & 0.912 & 0.880 \\
& \textbf{(0.007)} & (0.009) & \textbf{(0.008)} & \textbf{(0.011)} &
(0.020) & (0.050) & (0.021) & (0.092) \\

\hline

\multirow{2}{*}{\parbox{1.55cm}{Phys-Strat}} &
\textbf{0.970} & \textbf{0.969} & 0.958 & \textbf{0.957} &
\textbf{0.929} & \textbf{0.911} & \textbf{0.922} & 0.894 \\
& \textbf{(0.005)} & \textbf{(0.005)} & (0.004) & \textbf{(0.005)} &
\textbf{(0.015)} & \textbf{(0.011)} & \textbf{(0.021)} & (0.040) \\

\hline

\multirow{2}{*}{\parbox{3cm}{Phys-Strat-Aug}} &
\textbf{0.971} & \textbf{0.971} &  \textbf{0.962} & \textbf{0.960} & 
\textbf{0.930} & \textbf{0.913} & \textbf{0.921} & \textbf{0.899} \\
& \textbf{(0.004)} & \textbf{(0.005)} & \textbf{(0.003)} & \textbf{(0.004)} & 
\textbf{(0.016)} & \textbf{(0.019)} & \textbf{(0.015)} & \textbf{(0.019)} \\
\hline
\end{tabularx}
\end{table}

We extend our volumetric consistency analysis by analysing out of distribution (OoD) samples. In this instance they are defined as simulated images whose sequence parameters lie outside of the training range. This not only results in images of unfamiliar contrasts, but also unseen parameters that are fed into the physics branch of the network. If our method has truly attained a measure of sequence invariance then it should be expected that both segmentation quality and volume consistency are maintained as the network should be able to extrapolate from the provided values. For MPRAGE, the OoD range is extended to [100-2000] ms, while for SPGR, the TR is extended to [10-200] ms, TE is extended to [2-20] ms, and FA is extended to [5-90] degrees.

\begin{table}[t!]
\caption{Coefficients of variation (CoV) for Baseline, Phys-Base, Phys-Strat, and Phys-Strat-Aug on segmentation task, averaged across test subjects. Standard deviations quoted in brackets. Bold values represent statistically best performances.}\label{tab:tabCoV}
\centering
\begin{tabularx}{\linewidth}{|Y|v|v|v|v|v|v|v|v|}
\hline
\multirow{3}{*}{Experiments} & \multicolumn{8}{c|}{Sequence CoVs (x$10^3$)} \\ \cline{2-9}
& \multicolumn{4}{c|}{MPRAGE} & \multicolumn{4}{c|}{SPGR} \\ \cline{2-9}
& \multicolumn{2}{c|}{GM} & \multicolumn{2}{c|}{WM} & \multicolumn{2}{c|}{GM} & \multicolumn{2}{c|}{WM} \\ \cline{2-9}
& IoD & OoD & IoD & OoD & IoD & OoD & IoD & OoD \\
\hline

\multirow{2}{*}{\parbox{1.5cm}{Baseline}}
& 6.39 & 22.50 & 14.94 & 51.12  &
61.91 & 170.10 & 32.57 & 158.93 \\

& (0.87) & (4.08) & (1.71) & (7.11) &
(7.61) & (31.32) & (11.98) & (16.83) \\

\hline

\multirow{2}{*}{\parbox{1.5cm}{Phys-Base}}
& 2.72 & 14.67 & 3.28 & 28.10  &
77.22 & 127.22 & 20.77 & 264.80 \\

& (2.12) & (7.30) & (2.01) & (3.98) &
(34.44) & (18.61) & (9.35) & (8.52) \\

\hline

\multirow{2}{*}{\parbox{1.55cm}{Phys-Strat}}
& 0.71 & 6.15 & 0.53 & \textbf{3.67} &
21.83 & 59.78 & 8.60 & 59.19 \\

& (0.23) & (1.51) & (0.25) & \textbf{(1.34)} &
(0.83) & (13.31) & (0.64) & (11.25) \\

\hline

\multirow{2}{*}{\parbox{3cm}{Phys-Strat-Aug}}
& \textbf{0.42} & \textbf{4.74} & \textbf{0.51} & \textbf{3.65}
& \textbf{15.76} & \textbf{28.88} & \textbf{7.12} & \textbf{44.78} \\

& \textbf{(0.22)} & \textbf{(1.30)} & \textbf{(0.23)} & \textbf{(0.62)} &
\textbf{(1.18)} & \textbf{(9.74)} & \textbf{(0.45)} & \textbf{(4.22)} \\

\hline

\end{tabularx}
\end{table}

Table~\ref{tab:tabDice} and Table~\ref{tab:tabCoV} show Dice and CoV performances, respectively. We carry out signed-rank Wilcoxon tests to test for statistically significant improvements, and bold the best model (p-value $<$ 0.01). Tests are carried out on CoV and Dice scores independently of each other. In instances where models may outperform baselines but are not statistically significantly different from each other, we bold both.  We verify an incremental gain in CoV and Dice with each added feature, the most pronounced of which results from the addition of the stratification loss, in terms of both in and out of distribution CoVs. This is expected, as directly optimising for consistency across realisations of the same subject should more strongly enforce volume consistency, enhancing the physics invariance.

Phys-Strat-Aug boasts the best performance overall, significantly outperforming both Baseline and Phys-Base with regards to CoV. Compared to Phys-Strat, the differences are not always statistically better for MPRAGE, but are so for SPGR. With more parameters at play, an augmentation scheme should become more relevant, as sampling from the parameter space should lead to a greater extrapolating ability, as the network is no longer constrained to learn from a more discrete training set, and will experience more varied realisations.

\begin{figure}[t!]
\centering
\includegraphics[width=1.0\textwidth]{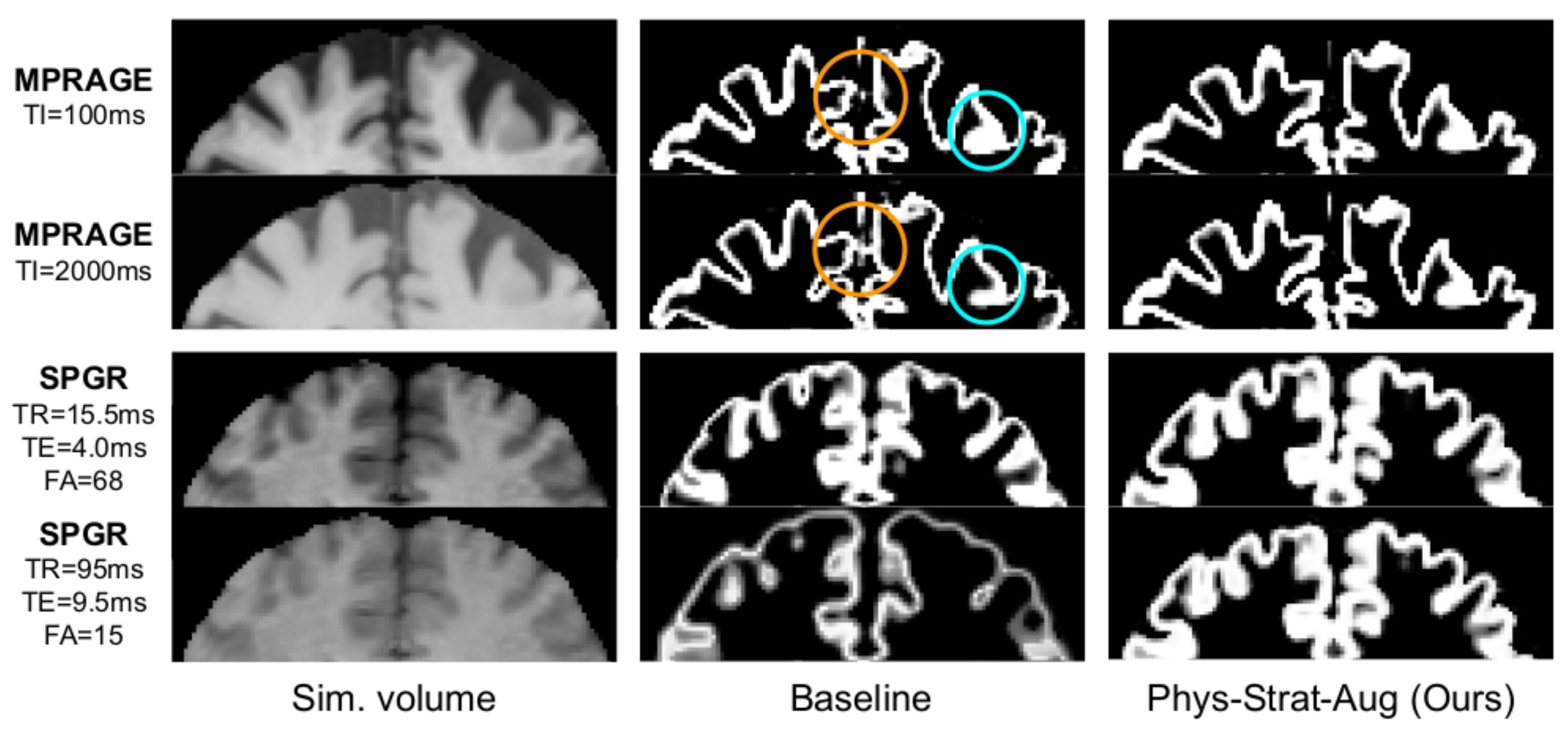} 
\caption{Baseline and Phys-Strat-Aug comparisons. Comparing out-of-distribution MPRAGE (Top two rows) and SPGR (Bottom two rows) GM segmentations from the proposed and baseline methods. Blue circles highlight examples of significant gyrus variability. Orange circles denote regions of segmentation differences between protocols.} 
\label{fig:figIoD_OoD_Comps}
\end{figure}

Fig.~\ref{fig:figIoD_OoD_Comps} shows some qualitative results, in and out of distribution segmentation comparisons between Baseline and Phys-Strat-Aug, to convey the consistency the latter is able to achieve without compromising segmentation quality.

\subsection{Uncertainty measures and volumetric bounds}

Given Phys-Strat-Aug's superior performance, we train only two epistemic uncertainty models, and two aleatoric uncertainty models with this pipeline, one of each for this pipeline and a complete baseline.

At test-time we extract 50 aleatoric volume samples, and 50 epistemic volume samples for each of the networks, for both in and out of distribution simulated images. We verify that the aleatoric samples do not contribute significantly to the volume variance in comparison to its epistemic counterpart, (an observation that was also been verified in~\cite{Eaton-Rosen2018}) and therefore omit it in our volumetric analysis. 

\begin{figure}[t!]
\centering
\includegraphics[width=1.0\textwidth]{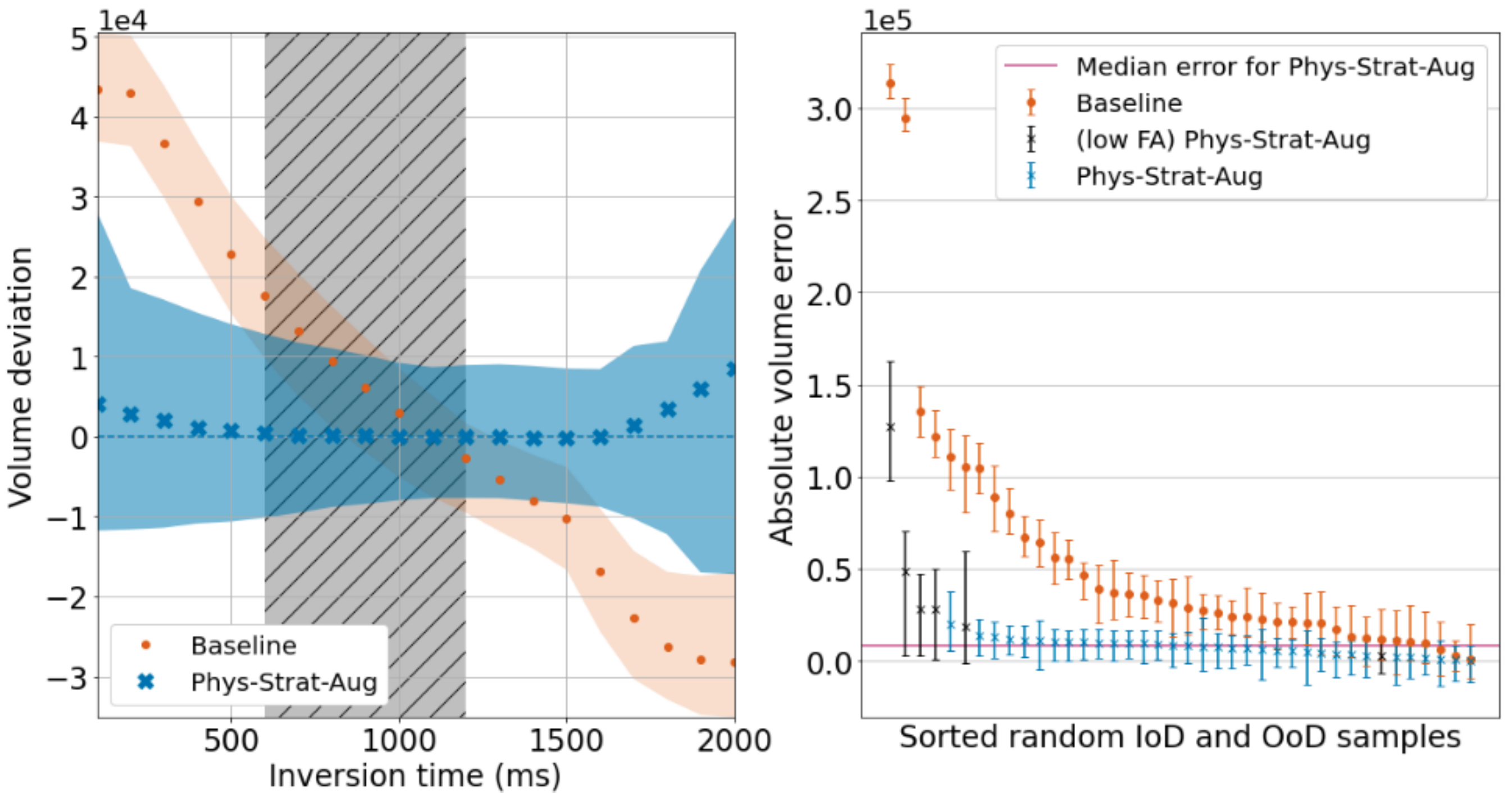} 
\caption{Volume consistency for WM for complete baseline and Phys-Strat-Aug, for example subject. Filled plots/ Error bars correspond to IQR volumes. Left: MPRAGE. The dashed grey region denotes the TI training time parameter range (600 - 1200 ms). Right: SPGR. Black points denote samples with FA lower than $10^{\circ}$ for Phys-Strat-Aug.} \label{fig:figMPRAGE_SPGR_Unc}
\end{figure}

Fig.~\ref{fig:figMPRAGE_SPGR_Unc} showcases white matter volume variations for MPRAGE and SPGR sequences, for the extended out of distribution parameter ranges, for Baseline and Phys-Strat-Aug experiments, for a single subject. For the SPGR plot, we order the points based on volumetric consistency for each experiment, thus highlighting outliers. In both instances we observe a much greater consistency in volume for Phys-Strat-Aug, itself a reflection of the aforementioned CoV results. Using the calibrated volumetric method described in~\cite{Eaton-Rosen2018} allows us to calculate volume percentiles for each set of dropout samples, and the errors represent the volumetric interquartile range (IQR). 

The errors for the baseline do not vary in any statistically significant manner, for either sequence or tissue, independent of any volume deviation. It is a different matter for Phys-Strat-Aug, however. Specifically, for MPRAGE, we note that uncertainties are consistently larger for Phys-Strat-Aug compared to baseline, and that furthermore, Phys-Aug-Strat segmentations boast larger uncertainties for out of distribution samples. This can perhaps be explained by the additional level of uncertainty introduced by the physics, and how the presence of a physics parameter outside of the ``known" further exacerbates this effect. 

For SPGR, all the apparent outliers for Phys-Strat-Aug have significantly larger associated errors, while this is not the case for the Baseline. We observe that most outliers correspond to out of distribution samples boasting very low flip angles ($<10^{\circ}$, highlighted in black in the figure). Such images will be significantly less $T_1$-weighted, and therefore be less familiar to the models, resulting in poorer segmentation quality, so it is reassuring that the physics-informed network's uncertainty around these samples is larger.

\section{Discussion and Conclusions}

In this work we demonstrated that with some well justified modifications to the training pipeline, a physics-informed network can achieve extremely constrained tissue segmentations across a wide range of contrasts, across all tissue types and investigated sequences; thus strengthening its harmonisation capabilities. 

Furthermore, we also showed that it can suitably generalise to unseen domains, while maintaining volume consistency without compromising segmentation quality, and is validated by accurately quantifying the volumetric uncertainty. The uncertainty estimates further suggest that the physics knowledge grants the model an additional level of safety, as volumetric uncertainties proved to be larger for out of distribution parameter generated images.

The method is admittedly limited by those sequences that can be aptly represented as a static equation, but we argue that at the very least, for the purposes of contrast agnosticism, a wide enough range of realistic contrasts can be generated with currently implemented sequences, which should allow for our method to generalise further. Future work will therefore involve testing of our method on multiple external datasets to ascertain generalisability and the exploration of techniques that may allow for the modelling of MR artifacts such as movement and $B_0$ inhomogeneities, to enhance our model's utility.

\subsubsection{Acknowledgements}
This project was funded by the Wellcome Flagship Programme (WT213038/Z/18/Z) and Wellcome EPSRC CME (WT203148/Z/16/Z).

\bibliographystyle{splncs04}
\bibliography{samplepaper}
\end{document}